\documentclass[10pt]{blois07}
\usepackage{graphicx}
\usepackage{cite,./mcite}

\usepackage{latexsym,amssymb,amsfonts,amsmath}

\def\be{\begin{equation}}
\def\ee{\end{equation}}
\def\bea{\begin{eqnarray}}
\def\eea{\end{eqnarray}}
\def\Tr{{\rm Tr}}
\def\re{{\rm Re}}
\def\im{{\rm Im}}

\begin{document}
\title{High--energy scattering and Euclidean--Minkowskian duality}
\author{Enrico Meggiolaro}
\institute{Dipartimento di Fisica, Universit\`a di Pisa, and INFN,
Sezione di Pisa,\\ Largo Pontecorvo 3, I--56127 Pisa, Italy.}
\maketitle

\begin{abstract}
We shall discuss how some relevant analyticity and crossing-symmetry
properties of the ``eikonal scattering amplitudes'' of two Wilson loops in QCD,
when going from Euclidean to Minkowskian theory, can be related to the still
unsolved problem of the asymptotic $s$--dependence of the hadron--hadron total
cross--sections. In particular, we critically discuss the question if (and how)
a {\it pomeron}--like behaviour can be derived from this Euclidean--Minkowskian
duality.
\end{abstract}

\section{Loop--loop and meson--meson scattering amplitudes}

\noindent
It was shown in Refs. \cite{DFK,Nachtmann97} (for a review see Refs.
\cite{Dosch,pomeron-book,LLCM1} and references therein)
that the high--energy meson--meson elastic
scattering amplitude can be approximately reconstructed in two steps:
i) one first evaluates, in the functional--integral approach, the high--energy
elastic scattering amplitude of two $q \bar{q}$ pairs (usually called
{\it dipoles}), of given transverse sizes $\vec{R}_{1\perp}$ and
$\vec{R}_{2\perp}$ and given longitudinal--momentum fractions $f_1$ and $f_2$
of the two quarks in the two dipoles respectively;
ii) one then averages this amplitude over all possible values of
$\vec{R}_{1\perp}, f_1$ and $\vec{R}_{2\perp}, f_2$ with two proper squared
wave functions $|\psi_1 (\vec{R}_{1\perp},f_1)|^2$ and
$|\psi_2 (\vec{R}_{2\perp},f_2)|^2$, describing the two interacting mesons.

The high--energy elastic scattering amplitude of two dipoles (defined in Eq.
(\ref{scatt-loop2}) below) is governed by the following (properly normalized)
connected correlation function of two Wilson loops forming an hyperbolic angle
$\chi$ in the longitudinal plane (see Eq. (\ref{p1p2}) below) and separated by
a distance $\vec{z}_\perp = (z^2,z^3)$ in the transverse plane
(impact parameter):
\be
{\cal C}_M(\chi,\vec{z}_\perp;1,2) \equiv
\displaystyle\lim_{T \to \infty}
\left[ { \langle {\cal W}^{(T)}_1 {\cal W}^{(T)}_2 \rangle \over
\langle {\cal W}^{(T)}_1 \rangle \langle {\cal W}^{(T)}_2 \rangle } -1 \right],
\label{C12}
\ee
where the arguments ``$1$'' and ``$2$'' in the function ${\cal C}_M$ stand for
``$\vec{R}_{1\perp}, f_1$'' and ``$\vec{R}_{2\perp}, f_2$'' respectively and
the expectation values $\langle \ldots \rangle$ are averages in the sense of
the QCD functional integrals. The two (infrared regularized) Wilson loops
${\cal W}^{(T)}_1$ and ${\cal W}^{(T)}_2$ are defined as:
\be
{\cal W}^{(T)}_{1,2} \equiv
{1 \over N_c} \Tr \left\{ {\cal P} \exp
\left[ -ig \displaystyle\oint_{{\cal C}_{1,2}} A_\mu(x) dx^\mu \right]
\right\} ,
\label{QCDloops}
\ee
where ${\cal C}_1$ and ${\cal C}_2$ are two rectangular paths which
follow the classical straight lines for quark [$X_{q}(\tau)$, forward in
proper time $\tau$] and antiquark [$X_{\bar{q}}(\tau)$, backward in $\tau$]
trajectories, i.e.,
\bea
{\cal C}_1 &:&
X_{1q}^\mu(\tau) = z^\mu + {p_1^\mu \over m} \tau + (1-f_1) R_1^\mu ,~~~~
X_{1\bar{q}}^\mu(\tau) = z^\mu + {p_1^\mu \over m} \tau - f_1 R_1^\mu ,
\nonumber \\
{\cal C}_2 &:&
X_{2q}^\mu(\tau) = {p_2^\mu \over m} \tau + (1-f_2) R_2^\mu ,~~~~
X_{2\bar{q}}^\mu(\tau) = {p_2^\mu \over m} \tau - f_2 R_2^\mu ,
\label{traj}
\eea
and are closed by straight--line paths at proper times $\tau = \pm T$, where
$T$ plays the role of an infrared cutoff, which can
and must be removed in the end ($T \to \infty$).
Here $p_1$ and $p_2$ are the four--momenta of the two dipoles, taken for
simplicity with the same mass $m$, moving (in the center--of--mass system)
with speed $V$ and $-V$ along, for example, the $x^1$--direction:
\be
p_1 =
m \left( \cosh {\chi \over 2},\sinh {\chi \over 2},\vec{0}_\perp \right) ,~~~
p_2 =
m \left( \cosh {\chi \over 2},-\sinh {\chi \over 2},\vec{0}_\perp \right) ,
\label{p1p2}
\ee
$\chi = 2~{\rm arctanh} V$ being the hyperbolic angle between the
two trajectories $1q$ and $2q$, i.e., $p_1 \cdot p_2 = m^2 \cosh\chi$.
Therefore, in terms of the usual Mandelstam variable $s$:
\be
s \equiv (p_1 + p_2)^2 = 2m^2 \left( \cosh\chi + 1 \right) ~,~~~~{\rm i.e.:}~~
\chi \mathop{\sim}_{s \to \infty} \log \left( {s \over m^2} \right) .
\label{logs}
\ee
It is convenient to consider also the correlation function
${\cal C}_E(\theta,\vec{z}_\perp;1,2)$ in the Euclidean
theory of two Euclidean Wilson loops running along two rectangular paths
$\widetilde{\cal C}_1$ and $\widetilde{\cal C}_2$, defined analogously to
(\ref{traj}), with the same $\vec{R}_{1\perp}$, $\vec{R}_{2\perp}$,
$\vec{z}_\perp$ and with the Minkowskian four--momenta $p_1$, $p_2$ replaced
by the following Euclidean four--vectors:
\be
p_{1E} =
m \left( \sin{\theta \over 2}, \vec{0}_\perp, \cos{\theta \over 2} \right) ,~~~
p_{2E} =
m \left( -\sin{\theta \over 2}, \vec{0}_\perp, \cos{\theta \over 2} \right) ,
\label{p1p2E}
\ee
$\theta$ being the angle formed by the two trajectories $1q$ and $2q$ in
Euclidean four--space, i.e., $p_{1E} \cdot p_{2E} = m^2 \cos\theta$.
It has been proved in Ref. \cite{Meggiolaro05} that
the Minkowskian quantity ${\cal C}_M$ with $\chi\in\mathbb{R}^+$ can be
reconstructed from the corresponding Euclidean quantity ${\cal C}_E$,
with $\theta \in (0,\pi)$, by an analytic continuation in the angular
variables $\theta \to -i\chi$, exactly as in the case of Wilson lines
\cite{Meggiolaro97,Meggiolaro98,Meggiolaro02}.
This result is derived under certain
hypotheses of analyticity in the angular variables \cite{crossing}.
In particular, one makes the assumption that
the function ${\cal C}_E$, as a function of the {\it complex}
variable $\theta$, can be {\it analytically extended} from the real segment
$(0 < \re\theta < \pi, \im\theta = 0)$ to a domain ${\cal D}_E$,
which also includes the negative imaginary axis
$(\re\theta = 0+, \im\theta < 0)$; and, therefore,
the function ${\cal C}_M$, as a function of the {\it complex} variable
$\chi$, can be {\it analytically extended} from the positive
real axis $(\re\chi > 0, \im\chi = 0+)$ to a domain
${\cal D}_M = \{ \chi \in \mathbb{C} ~|~ -i\chi \in {\cal D}_E \}$,
which also includes the imaginary segment $(\re\chi = 0, 0 < \im\chi < \pi)$.
The validity of this assumption is confirmed by explicit calculations in
perturbation theory \cite{Meggiolaro05,Meggiolaro97,BB}.
Denoting with $\overline{\cal C}_M$ and $\overline{\cal C}_E$ such analytic
extensions, we then have the following {\it analytic--continuation relations}
\cite{Meggiolaro05,crossing}:
\bea
\overline{\cal C}_E(\theta,\vec{z}_\perp;1,2) &=&
\overline{\cal C}_M(i\theta,\vec{z}_\perp;1,2) ,
\qquad \forall\theta\in {\cal D}_E ;
\nonumber \\
\overline{\cal C}_M(\chi,\vec{z}_\perp;1,2) &=&
\overline{\cal C}_E(-i\chi,\vec{z}_\perp;1,2) ,
\qquad \forall\chi\in {\cal D}_M .
\label{final}
\eea
The validity of the relation (\ref{final}) for the loop--loop correlators 
in QCD has been also recently verified in Ref. \cite{BB} by an explicit
calculation up to the order ${\cal O}(g^6)$ in perturbation theory.
However we want to stress that the analytic continuation (\ref{final})
is expected to be an {\it exact} result, i.e., not restricted
to some order in perturbation theory or to some other approximation,
and is valid both for the Abelian and the non--Abelian case.

The relation (\ref{final}) allows the derivation of the {\it loop--loop
scattering amplitude}, which is defined as
\be
{\cal M}_{(ll)} (s,t;\vec{R}_{1\perp},f_1,\vec{R}_{2\perp},f_2) = -i~2s~
\widetilde{\cal C}_M \left( \chi \mathop{\sim}_{s \to \infty}
\log \left( {s \over m^2} \right), t;1,2 \right) ,
\label{scatt-loop2}
\ee
$\widetilde{\cal C}_M$ being the two--dimensional Fourier transform of
${\cal C}_M$, with respect to the impact parameter $\vec{z}_\perp$,
at transferred momentum $\vec{q}_\perp$ (with $t = -|\vec{q}_\perp|^2$), i.e.,
\be
\widetilde{\cal C}_M (\chi, t;1,2) \equiv
\displaystyle\int d^2 \vec{z}_\perp e^{i \vec{q}_\perp \cdot \vec{z}_\perp}
{\cal C}_M (\chi,\vec{z}_\perp;1,2) ,
\label{CMtilde}
\ee
from the analytic continuation $\theta \to -i\chi$ of the corresponding
Euclidean quantity:
\be
\widetilde{\cal C}_E (\theta, t;1,2) \equiv
\displaystyle\int d^2 \vec{z}_\perp e^{i \vec{q}_\perp \cdot \vec{z}_\perp}
{\cal C}_E (\theta,\vec{z}_\perp;1,2) ,
\label{CEtilde}
\ee
which can be evaluated non-perturbatively by well--known and well--established
techniques available in the Euclidean theory.
This approach has been extensively used in the literature
\cite{LLCM2,instanton1,instanton2,JP2,Janik} in order to
tackle, from a theoretical point of view, the still unsolved problem
of the asymptotic $s$--dependence of hadron--hadron elastic scattering
amplitudes and total cross sections.
As we have already said in the beginning, the {\it hadron--hadron elastic
scattering amplitude} ${\cal M}_{(hh)}$ can be obtained by averaging the
loop--loop scattering amplitude (\ref{scatt-loop2}) over all possible dipole
transverse separations $\vec{R}_{1\perp}$ and $\vec{R}_{2\perp}$ and
longitudinal--momentum fractions $f_1$ and $f_2$ with two proper squared
hadron wave functions \cite{DFK,Nachtmann97,Dosch,pomeron-book,LLCM1}:
\bea
{\cal M}_{(hh)}(s,t) &=&
\displaystyle\int d^2\vec{R}_{1\perp} \int_0^1 df_1~
|\psi_1(\vec{R}_{1\perp},f_1)|^2
\displaystyle\int d^2\vec{R}_{2\perp} \int_0^1 df_2~
|\psi_2(\vec{R}_{2\perp},f_2)|^2
\nonumber \\
&\times& {\cal M}_{(ll)} (s,t;\vec{R}_{1\perp},f_1,\vec{R}_{2\perp},f_2) .
\label{scatt-hadron}
\eea
Denoting with ${\cal C}_M^{(hh)}$ and ${\cal C}_E^{(hh)}$ the quantities
obtained by averaging the corresponding loop--loop correlation functions
${\cal C}_M$ and ${\cal C}_E$ over all possible dipole transverse separations
$\vec{R}_{1\perp}$ and $\vec{R}_{2\perp}$ and longitudinal--momentum fractions
$f_1$ and $f_2$, in the same sense as in Eq. (\ref{scatt-hadron}),
we can write:
\be
{\cal M}_{(hh)} (s,t) = -i~2s~
\widetilde{\cal C}_M^{(hh)} \left( \chi \mathop{\sim}_{s \to \infty}
\log \left( {s \over m^2} \right), t \right) .
\label{scatt-hadron2}
\ee
Clearly, by virtue of the relation (\ref{final}), we also have that:
\be
\overline{\widetilde{\cal C}_M^{(hh)}} (\chi, t) =
\overline{\widetilde{\cal C}_E^{(hh)}} (-i\chi, t) ,
\qquad \forall\chi\in {\cal D}_M .
\label{final-hh}
\ee
By virtue of the optical theorem, the hadron--hadron total cross section can be
derived from the imaginary part of the forward hadron--hadron elastic
scattering amplitude.
Experimental observations at the present time seem to be well described by a
{\it pomeron}--like high--energy behaviour (see, for example, Ref.
\cite{pomeron-book} and references therein):
\be
\sigma_{\rm tot}^{(hh)} (s) \mathop{\sim}_{s \to \infty}
{1 \over s} {\rm Im} {\cal M}_{(hh)} (s, t=0) \sim
\sigma_0^{(hh)} \left( {s \over s_0} \right)^{\epsilon_P} ,
~~~~{\rm with}~~\epsilon_P \simeq 0.08 .
\label{pomeron}
\ee
A behaviour like the one of Eq. (\ref{pomeron}) seems to emerge directly
(apart from possible undetermined $\log s$ prefactors) when applying
the Euclidean--to--Minkowskian analytic--continuation approach to the
study of the line--line/loop--loop scattering amplitudes in strongly coupled
(confining) gauge theories using the AdS/CFT correspondence \cite{JP2,Janik}.

Moreover, it has been found in Ref. \cite{BB} that
the dipole--dipole cross section, evaluated from the loop--loop correlator
up to the order ${\cal O}(g^6)$, reproduces the first iteration of the BFKL
{\it kernel} in the leading--log approximation, the so--called
BFKL--{\it pomeron} behaviour, i.e., $\sim s^{{12\alpha_s \over \pi}\log 2}$,
with $\alpha_s = g^2/4\pi$ \cite{BFKL}.

\section{How a pomeron--like behaviour can be derived}

\noindent
The way in which a {\it pomeron}--like behaviour can emerge, using the
Euclidean--to--Minkowskian analytic continuation, was first shown in Ref.
\cite{Meggiolaro97} in the case of the line--line (i.e., parton--parton)
scattering amplitudes. Here we shall readapt that analysis to the case of the
loop--loop scattering amplitudes, with more technical developments, new
interesting insights and critical considerations \cite{Meggiolaro07}.
We start by writing the Euclidean hadronic correlation function
in a partial--wave expansion:
\be
\widetilde{\cal C}_E^{(hh)} (\theta,t) =
\displaystyle\sum_{l=0}^{\infty} (2l+1) A_l(t) P_l (\cos \theta) .
\label{pwe}
\ee
As shown in Ref. \cite{crossing},
the loop--antiloop correlator at angle $\theta$ in the
Euclidean theory (or at hyperbolic angle $\chi$ in the Minkowskian theory) can
be derived from the corresponding loop--loop correlator by the substitution
$\theta \to \pi - \theta$ (or $\chi \to i\pi - \chi$ in the Minkowskian
theory). Because of these {\it crossing--symmetry relations}, it is natural
to decompose also our hadronic correlation function
$\widetilde{\cal C}_E^{(hh)} (\theta,t)$ as a sum of a
{\it crossing--symmetric} function $\widetilde{\cal C}_E^+ (\theta,t)$
and of a {\it crossing--antisymmetric} function
$\widetilde{\cal C}_E^- (\theta,t)$:
\be
\widetilde{\cal C}_E^{(hh)} (\theta,t) =
\widetilde{\cal C}_E^+ (\theta,t) + \widetilde{\cal C}_E^- (\theta,t) ,~~~~
\widetilde{\cal C}_E^{\pm} (\theta,t) \equiv
{\widetilde{\cal C}_E^{(hh)} (\theta,t) \pm \widetilde{\cal C}_E^{(hh)}
(\pi-\theta,t) \over 2} .
\ee
Using Eq. (\ref{pwe}), we can find the partial--wave expansions of these two
functions as follows:
\be
\widetilde{\cal C}_E^{\pm} (\theta,t) = {1 \over 2}
\displaystyle\sum_{l=0}^{\infty} (2l+1) A_l(t)
[P_l (\cos \theta) \pm P_l (-\cos \theta)] .
\label{pwe-pm}
\ee
Because of the relation $P_l(-\cos \theta) = (-1)^l P_l(\cos \theta)$, valid
for non--negative integer values of $l$, we immediately see that
$\widetilde{\cal C}_E^+ (\theta,t)$ gets contributions only from even
$l$, while $\widetilde{\cal C}_E^- (\theta,t)$ gets contributions only from
odd $l$. For this reason the functions
$\widetilde{\cal C}_E^{\pm} (\theta,t)$ can also be called
{\it even--signatured} and {\it odd--signatured} correlation functions
respectively and we can replace $A_l(t)$ in Eq. (\ref{pwe-pm}) respectively
with $A_l^\pm (t) \equiv {1 \over 2} [1 \pm (-1)^l] A_l(t)$.
However, if we write the hadronic correlation function
$\widetilde{\cal C}_E^{(hh)} (\theta,t)$
in terms of the loop--loop correlation function, averaged
over all possible dipole transverse separations $\vec{R}_{1\perp}$ and
$\vec{R}_{2\perp}$ and longitudinal--momentum fractions $f_1$ and $f_2$ with
two proper squared hadron wave functions 
$|\psi_1(\vec{R}_{1\perp},f_1)|^2$ and $|\psi_2(\vec{R}_{2\perp},f_2)|^2$,
and we make use:
i) of the so--called {\it crossing--symmetry relations} for loop--loop
correlators \cite{crossing}:
\bea
\lefteqn{
\mathcal{C}_E(\pi-\theta,\vec{z}_{\perp};\vec{R}_{1\perp},f_1,
\vec{R}_{2\perp},f_2) }
\label{crossingC} \\
& & =\mathcal{C}_E(\theta,\vec{z}_{\perp};\vec{R}_{1\perp},f_1,
-\vec{R}_{2\perp},1-f_2)
=\mathcal{C}_E(\theta,\vec{z}_{\perp};-\vec{R}_{1\perp},1-f_1,
\vec{R}_{2\perp},f_2) ,
\quad\forall\theta\in\mathbb{R} ;
\nonumber
\eea
and ii) of the rotational-- and $C$--invariance of the squared hadron wave
functions, that is:
\be
|\psi_i(\vec{R}_{i\perp},f_i)|^2 = |\psi_i(-\vec{R}_{i\perp},f_i)|^2 =
|\psi_i(\vec{R}_{i\perp},1-f_i)|^2 = |\psi_i(-\vec{R}_{i\perp},1-f_i)|^2
\ee
(see Refs. \cite{Dosch,LLCM1} and also \cite{pomeron-book}, chapter 8.6, and
references therein), then we immediately conclude that the hadronic correlation
function $\widetilde{\cal C}_E^{(hh)} (\theta,t)$ is automatically
crossing symmetric and so it coincides with the even--signatured
function $\widetilde{\cal C}_E^+ (\theta,t)$, the odd--signatured
function $\widetilde{\cal C}_E^- (\theta,t)$ being identically equal to zero.
Upon analytic continuation from the Euclidean to the Minkowskian theory (see
again Ref. \cite{crossing}), this means that the Minkowskian hadronic
correlation function
$\widetilde{\cal C}_M^{(hh)} (\chi,t)$, and therefore also the scattering
amplitude ${\cal M}_{(hh)}$ written in Eq. (\ref{scatt-hadron2}), turns out to
be automatically crossing symmetric, i.e., invariant under the exchange
$\chi \to i\pi - \chi$:
$\widetilde{\cal C}_M^{(hh)} (\chi,t) = \widetilde{\cal C}_M^+ (\chi,t)$,
$\widetilde{\cal C}_M^- (\chi,t) = 0$.
In other words, our formalism naturally leads to a high--energy meson--meson
scattering amplitude which, being crossing symmetric, automatically
satisfies the Pomeranchuk theorem. An {\it odderon} (i.e., $C=-1$) exchange
seems to be excluded for high--energy meson--meson scattering, while a
{\it pomeron} (i.e., $C=+1$) exchange is possible \cite{odderon}.

Let us therefore proceed by considering our {\it crossing--symmetric}
Euclidean correlation function:
\be
\widetilde{\cal C}_E^{(hh)} (\theta,t) =
\widetilde{\cal C}_E^+ (\theta,t) = {1 \over 2}
\displaystyle\sum_{l=0}^{\infty} (2l+1) A_l^+ (t)
[P_l (\cos \theta) + P_l (-\cos \theta)] .
\label{pwe-E}
\ee
We can now use Cauchy's theorem to rewrite this partial--wave expansion as an
integral over $l$, the so--called {\it Sommerfeld--Watson transform}:
\be
\widetilde{\cal C}_E^{(hh)} (\theta,t) = \widetilde{\cal C}_E^+ (\theta,t) =
-{1 \over 4i} \displaystyle\int_C {(2l+1) A_l^+(t)
[ P_l(-\cos\theta) + P_l(\cos\theta)] \over \sin(\pi l)} dl ,
\label{swt}
\ee
where ``$C$'' is a contour in the complex $l$--plane, running clockwise
around the real positive $l$--axis and enclosing all non--negative integers,
while excluding all the singularities of $A_l^+$.
Here (as in the original derivation: see, e.g., Ref. \cite{pomeron-book} and
references therein) we make the fundamental {\it assumption} that the
singularities of $A_l^+(t)$ in the complex $l$--plane (at a given $t$)
are only {\it simple poles}.
(However, we want to remark that our {\it partial--wave amplitudes} $A_l^+(t)$
are {\it not} the same partial--wave amplitudes considered in the
original derivation.)
Then we can use again Cauchy's theorem to reshape the contour $C$ into the
straight line $\re (l) = -{1 \over 2}$ and rewrite the integral (\ref{swt})
as follows:
\bea
\lefteqn{
\widetilde{\cal C}_E^{(hh)} (\theta,t) = \widetilde{\cal C}_E^+ (\theta,t) = }
\nonumber \\
& & -{\pi \over 2} \displaystyle\sum_{ \re (\sigma_n^+) >  -{1 \over 2} }
{ (2\sigma_n^+ (t) + 1) r_n^+ (t) [P_{\sigma_n^+ (t)} (-\cos \theta)
+ P_{\sigma_n^+ (t)}(\cos \theta)] \over \sin (\pi \sigma_n^+ (t))}
\nonumber \\
& & -{1 \over 4i}
\displaystyle\int_{-{1 \over 2}-i\infty}^{-{1 \over 2}+i\infty}
{ (2l+1) A_l^+ (t) [P_l (-\cos \theta) + P_l (\cos \theta)] \over
\sin (\pi l) } dl ,
\label{swt-E}
\eea
where $\sigma_n^+ (t)$ is a pole of $A_l^+ (t)$ in the complex $l$--plane and
$r_n^+ (t)$ is the corresponding residue. We have also assumed that the
large--$l$ behaviour of $A_l^+$ is such that the integrand function in
Eq. (\ref{swt}) vanishes enough rapidly (faster than
$1/l$) as $|l| \to \infty$ in the right half--plane, so that the 
contribution from the infinite contour is zero.

Eq. (\ref{swt-E}) immediately
leads to the asymptotic behaviour of the scattering amplitude in the limit
$s \to \infty$, with a fixed $t$ ($|t| \ll s$). In fact, making use of the
analytic extension (\ref{final-hh}) when continuing the angular variable,
$\theta \to -i\chi$, we derive that for every $\chi\in\mathbb{R}^+$:
\bea
\lefteqn{
\widetilde{\cal C}_M^{(hh)} (\chi, t) =
\overline{\widetilde{\cal C}_E^{(hh)}} (-i\chi, t) = } \nonumber \\
& & -{\pi \over 2} \displaystyle\sum_{ \re (\sigma_n^+) >  -{1 \over 2} }
{ (2\sigma_n^+ (t) + 1) r_n^+ (t) [P_{\sigma_n^+ (t)} (-\cosh \chi)
+ P_{\sigma_n^+ (t)}(\cosh \chi)] \over \sin (\pi \sigma_n^+ (t))}
\nonumber \\
& & -{1 \over 4i}
\displaystyle\int_{-{1 \over 2}-i\infty}^{-{1 \over 2}+i\infty}
{ (2l+1) A_l^+ (t) [P_l (-\cosh \chi) + P_l (\cosh \chi)] \over
\sin (\pi l) } dl .
\label{swt-M}
\eea
Now we must take the large--$\chi$ (large--$s$) limit of this expression, with
the hyperbolic angle $\chi$ expressed in terms of $s$ by the relation
(\ref{logs}), i.e., $\cosh \chi = {s \over 2m^2} - 1$.
The asymptotic form of $P_\nu (z)$ when $z \to \infty$ is known to be
a linear combination of $z^\nu$ and of $z^{-\nu -1}$.
When $\re (\nu) >  -1/2$, the last term can be neglected and thus,
in the limit $s \to \infty$, with a fixed $t$ ($|t| \ll s$), we obtain, from
the sum in Eq. (\ref{swt-M}) (see Ref. \cite{Meggiolaro07} for more details):
\be
\widetilde{\cal C}_M^{(hh)} \left( \chi \mathop{\sim}_{s \to \infty}
\log \left( {s \over m^2} \right), t \right)
\sim \displaystyle\sum_{ \re (\sigma_n^+) >  -{1 \over 2} }
\beta_n^+ (t) s^{\sigma_n^+ (t)} .
\label{CMhh-regge}
\ee
The integral in Eq. (\ref{swt-M}), usually called the {\it background term},
vanishes at least as $1/\sqrt{s}$ and therefore can be neglected.
From eqs. (\ref{scatt-hadron2}) and (\ref{CMhh-regge}) we can extract the
elastic scattering amplitude:
\be
{\cal M}_{(hh)} (s,t) \mathop{\sim}_{s \to \infty}
-2i \displaystyle\sum_{ \re (\sigma_n^+) > -{1 \over 2} }
\beta_n^+ (t) s^{1+\sigma_n^+ (t)} .
\label{Mhh-regge}
\ee
This equation gives the explicit $s$--dependence of the scattering amplitude at 
very high energy ($s \to \infty$) and small transferred momentum ($|t| \ll s$).
As we can see, this amplitude comes out to be a sum of powers of $s$.
This sort of behaviour for the scattering amplitude is known in the literature
as a {\it Regge behaviour} and $1+\sigma_n^+ (t) \equiv \alpha_n^+ (t)$ is
the so--called {\it Regge trajectory}.
In the original derivation (see, e.g., Ref. \cite{pomeron-book} and
references therein) the asymptotic
behaviour (\ref{Mhh-regge}) is recovered by analytically continuing the
$t$--channel scattering amplitude to very large imaginary values of the angle
between the trajectories of the two exiting particles in the $t$--channel
scattering process.
Instead, in our derivation (see Ref. \cite{Meggiolaro07}),
we have used the Euclidean--to--Minkowskian
analytic continuation (\ref{final-hh}) and we have analytically continued
the Euclidean loop--loop correlator to very large (negative) imaginary values 
of the angle $\theta$ between the two Euclidean Wilson loops.

Denoting with $\sigma_P (t)$ the pole with the largest real part (at
that given $t$) and with $\beta_P (t)$ the corresponding coefficient
$\beta_n^+ (t)$ in Eq. (\ref{CMhh-regge}), we thus find that:
\be
\widetilde{\cal C}_M^{(hh)} \left( \chi \mathop{\sim}_{s \to \infty}
\log \left( {s \over m^2} \right), t \right)
\sim \beta_P (t) s^{\sigma_P (t)} \Longrightarrow
{\cal M}_{(hh)} (s,t) \mathop{\sim}_{s \to \infty}
-2i~\beta_P (t)~ s^{\alpha_P (t)} ,
\label{Mhh-asympt}
\ee
where $\alpha_P (t) \equiv 1 + \sigma_P (t)$ is the {\it pomeron trajectory}.
Therefore, by virtue of the optical theorem:
\be
\sigma_{\rm tot}^{(hh)} (s) \mathop{\sim}_{s \to \infty}
{1 \over s} {\rm Im} {\cal M}_{(hh)} (s, t=0)
\sim \sigma_0^{(hh)} \left( {s \over s_0} \right)^{\epsilon_P} ,
~~~~{\rm with}~~\epsilon_P = \re[\alpha_P (0)]-1 .
\label{pomeron2}
\ee
We want to stress two important issues which clarify under which conditions
we have been able to derive this {\it pomeron}--like behaviour for the
elastic amplitudes and the total cross sections.

{\bf i)} We have ignored a possible energy dependence of hadron wave functions
and we have thus ascribed the high--energy behaviour of the Minkowskian
hadronic correlation function exclusively to the {\it fundamental} loop--loop
correlation function (\ref{CMtilde}). With this hypothesis, the coefficients
$A_l^+$ in the partial--wave expansion (\ref{pwe}) and, as a consequence, the
coefficients $\beta_n^+$ and $\sigma_n^+$ in the Regge expansion
(\ref{CMhh-regge}) do not depend on $s$, but they only depend on the
variable $t$.

{\bf ii)} However, this is not enough to guarantee the experimentally--observed
{\it universality} (i.e., independence on the specific type of hadrons
involved in the reaction) of the {\it pomeron} trajectory
$\alpha_P (t)$ in Eq. (\ref{Mhh-asympt}) and, therefore, of the
{\it pomeron} intercept $1+\epsilon_P$ in Eq. (\ref{pomeron2}).
In fact, the partial--wave expansion (\ref{pwe}) of the hadronic correlation
function can also be considered as a result of a partial--wave
expansion of the loop--loop Euclidean correlation function
(\ref{CEtilde}), i.e.,
\be
\widetilde{\cal C}_E (\theta,t;1,2) =
\displaystyle\sum_{l=0}^{\infty} (2l+1) {\cal A}_l(t;1,2) P_l (\cos \theta) ,
\label{CEtilde-pwe}
\ee
which is then averaged with two proper squared hadron wave functions,
in the same sense as in Eq. (\ref{scatt-hadron}), so giving the Euclidean
hadronic correlation function (\ref{pwe}).
If we now repeat for the partial--wave expansion (\ref{CEtilde-pwe}) the
same manipulations that have led us from Eq. (\ref{pwe}) to Eq.
(\ref{CMhh-regge}), we arrive at the following Regge expansion for the
(even--signatured) loop--loop Minkowskian correlator:
\be
\widetilde{\cal C}_M^+ \left( \chi \mathop{\sim}_{s \to \infty}
\log \left( {s \over m^2} \right), t;1,2 \right)
\sim \displaystyle\sum_{ \re (a_n^+) >  -{1 \over 2} }
b_n^+ (t;1,2) s^{a_n^+ (t;1,2)} ,
\label{CM-regge}
\ee
where $a_n^+ (t;1,2)$ is a pole of ${\cal A}_l^+ (t;1,2)$ in the complex
$l$--plane. After inserting the expansion (\ref{CM-regge}) into the expression
for the Minkowskian hadronic correlation function:
\bea
\widetilde{\cal C}_M^{(hh)} (\chi,t) &=&
\displaystyle\int d^2\vec{R}_{1\perp} \int_0^1 df_1~
|\psi_1(\vec{R}_{1\perp},f_1)|^2
\displaystyle\int d^2\vec{R}_{2\perp} \int_0^1 df_2~
|\psi_2(\vec{R}_{2\perp},f_2)|^2
\nonumber \\
&\times& \widetilde{\cal C}_M^+ (\chi,t;1,2) ,
\label{CMhh-tilde}
\eea
one in general finds a high--energy behaviour which hardly fits with that
reported in Eq. (\ref{Mhh-asympt}) with a
universal {\it pomeron} trajectory $\alpha_P (t)$, {\it unless}
one assumes that, for each given loop--loop correlation function with
transverse separations $\vec{R}_{1\perp}$ and $\vec{R}_{2\perp}$ and
longitudinal--momentum fractions $f_1$ and $f_2$, (at least)
the location of the pole $a_n^+ (t;1,2)$ with the largest real part does not
depend on $\vec{R}_{1\perp},f_1$ and $\vec{R}_{2\perp},f_2$,
but only depends on $t$.
If we denote this common pole with $\sigma_P (t)$,
we then immediately recover the high--energy behaviour
(\ref{Mhh-asympt}), where the coefficient $\beta_P (t)$ in front,
differently from the universal function $\alpha_P (t) = 1 + \sigma_P (t)$,
explicitly depends on the specific type of hadrons involved in the process.

\section{Conclusions and outlook}

\noindent
In conclusion, we have shown that the Euclidean--to--Minkowskian
analytic--continuation approach can, with the inclusion of some extra
(more or less plausible) {\it assumptions}, easily reproduce a
{\it pomeron}--like behaviour for the high--energy total cross sections,
in apparent agreement with the present--time experimental observations.
However, we should also keep in mind that the {\it pomeron}--like
behaviour (\ref{pomeron}) is, strictly speaking, theoretically forbidden (at
least if considered as a true {\it asymptotic} behaviour) by the well--known
Froissart--Lukaszuk--Martin (FLM) theorem \cite{FLM}.
In this respect, the {\it pomeron}--like behaviour (\ref{pomeron}) can at most
be regarded as a sort of {\it pre--asymptotic} (but not really
{\it asymptotic}) behaviour of the high--energy total cross sections, valid
in a certain high--energy range.\\
Immediately the following question arises: why our approach, which was
formulated so to give the really asymptotic large--$s$ behaviour of
scattering amplitudes and total cross sections, is also able to reproduce
pre--asymptotic behaviours (violating the FLM bound) like the
one in (\ref{pomeron})?\\
The answer is clearly that the extra {\it assumptions}, i.e., the {\it models},
which one implicitly or explicitly uses in the calculation of the
Euclidean correlation function $\widetilde{\cal C}_E$, play a fundamental
role in this respect.
Of course, every model has its own {\it limitations}, which reflect in the
variety of answers in the literature.
Unfortunately these {\it limitations} are often out of control, in the sense
that no one knows exactly what is losing due to these approximations.
This is surely a crucial point which, in our opinion, should be further
investigated in the future, also with the help of direct lattice calculations
of the loop--loop Euclidean correlation function.

%------------------------------------------------------------------------------
%       Bibliography
%------------------------------------------------------------------------------
\begin{footnotesize}
\bibliographystyle{blois07} 
{\raggedright
\bibliography{blois07}
}

\end{footnotesize}

\end{document}